\documentclass[]{aastex631}

\shorttitle{Plaskett Observation of Starlink}
\shortauthors{Boley et al.}
\graphicspath{{./}{figures/}}
\newcommand{\Gaia}{{\it Gaia}}

\begin{document}

\title{Plaskett 1.8 metre Observations of Starlink Satellites}

\author[0000-0002-0574-4418]{Aaron C.~Boley}
\affiliation{Department of Physics and Astronomy, University of British Columbia \\
6224 Agricultural Rd\\
Vancouver, BC V6T 1Z1, Canada}

\author{Ewan Wright}
\affiliation{ISGP, University of British Columbia \\
2357 Main Mall\\
Vancouver, BC V6T 1Z1, Canada}

\author{Samantha Lawler}
\affiliation{Campion College and the Department of Physics, University of Regina\\
3737 Wascana Parkway, Regina, SK S4S 0A2}

\author{Paul Hickson}
\affiliation{Department of Physics and Astronomy, University of British Columbia \\
6224 Agricultural Rd\\
Vancouver, BC V6T 1Z1, Canada}

\author{Dave Balam}
\affiliation{Dominion Astrophysical Observatory \\
5071 West Saanich Road\\
Victoria, BC V9E 2E7, Canada}

\begin{abstract}

We present observations of 23 Starlink satellites in the $g'$ bandpass, obtained from the Dominion Astrophysical Observatory's Plaskett 1.8 metre telescope. The targets include a mixture of satellites with and without brightness mitigation measures (i.e., visors). 
At the time of the observations (16 July 2021), Starlink satellites were sunlight throughout the night, and even with strict elevation and azimuth limits, there were over 800 candidate Starlink arcs. 
The satellites altogether have a median absolute brightness (550 km) of $\overline{H}_g^{550} =5.3$ mag. 
Dividing the targets into those without and with visors, their median absolute magnitudes are $\overline{H}_g^{550}(no~visor)=5.1$ and $\overline{H}^{550}_g(visor)=5.7$ mag, respectively. 
While the visor sample is dimmer in aggregate, the absolute brightness distribution ranged from $H_g^{550}=4.3$ mag to 9.4 mag, with the brightest being a visored satellite and the dimmest a satellite with no mitigation.  
The intrinsic brightness dispersion among the full sample is  $\sigma_g = 0.5$ mag.

\end{abstract}


\section{Introduction} \label{sec:intro}

Large constellations of satellites will have a substantial impact on astronomy, particularly wide-field surveys, should they be deployed in their proposed numbers. 
These so-called `mega-constellations' or `satcons' seek to place tens of thousands of communications satellites into low-Earth orbit (LEO), with the total number of proposed satellites exceeding 100,000. 
Starlink \citep{FCCStarlinkInitial,FCCStarlinkVLEO,FCCStarlinkGen1Mod,FCCStarlinkGen2}, OneWeb \citep{OneWebPhase2}, Amazon/Kuiper \citep{Kuiper}, and GW/StarNet \citep{press2021} alone could place approximately 65,000 satellites into orbit, although this number is being continuously amended. 
While not all of the proposed satellites are expected to be deployed, other operators are entering the market, making a plausible situation in which LEO will have 100,000 or more satellites should no restrictions be put into place. 

There are ongoing efforts by astronomers to address the potential for NewSpace satellite light pollution to interfere with astronomy, including engaging in dialogue with satcon operators (e.g., SATCON2\footnote{\url{https://noirlab.edu/satcon2/}} and Dark and Quiet Skies\footnote{https://www.iau.org/news/announcements/detail/ann21043/}). 
Such dialogue seeks to provide opportunities for exchanges of views and to establish a basis for cooperation, with identifying voluntary mitigation efforts for the short term and defining feasible and effective regulations or guidelines for the protection of the night sky in the long-term. 
As part of this effort, it is incumbent upon astronomers to routinely measure satellites for the purpose of providing independent brightness assessments in astronomical observing bands and to verify mitigation practices. 

Several groups have reported on LEO satellite magnitudes under various conditions \citep{TysonLSST,tregloan-reed2021,Mallama2021}.
This includes efforts to focus on specific satellites to assess mitigation attempts, such as `DarkSat' and `Visorsats', representing albedo and reflection modifications, respectively. 

We contribute to this effort by observing 23 Starlink satellites using the Plaskett 1.8 metre telescope at the Dominion Astrophysical Observatory. The sample includes a mixture of unmitigated Starlinks and visored Starlinks, which we analyze together and separately.

\section{Target Selection and Observing Method}

The Two-Line Elements (TLEs) for all Starlink satellites were obtained from Celestrak\footnote{\url{https://celestrak.com/}} on July 15, 2021 UTC. 
Each element was then run through the JPL Horizons ephemeris service\footnote{\url{https://ssd.jpl.nasa.gov/horizons.cgi}} using an API to determine the observing opportunities on July 16, 2021 UTC. 
An initial selection was made for Starlink satellites that (1) were visible from DAO during the observing night, including astronomical twilight, (2) reached a maximum elevation that was greater than $45^\circ$, and (3) reached their maximum elevation in the southern sky, with this latter being mainly a telescope slewing consideration. 
{\it This resulted in over 800 observing opportunities during the approximately three hours of useful night.}

Refining the targets further was motivated largely by the satellite observing method.  
The Plaskett 1.8 metre is unable to track LEO objects.
Instead, a `wait and catch' methodology is used. 
In other words, the telescope points at a field, and tracks it at the sidereal rate. 
An exposure is then made such that the satellite of interest is expected to pass entirely through the telescope's field of view (FOV) during the image integration. 

One of the largest uncertainties in the observing plan is the time needed to slew between all of the different targets. 
A conservative approach was taken, and altogether, we required that the spacing between observations was 12 minutes or more -- carrying out this program demonstrated that a slightly faster cadence would be possible for similar future observing plans. 
In the end, the time between the planned observations varied between 12 and 15 minutes, allowing the selection of 23 targets, all of which were successfully observed. 

Among the sample, nine are unmitigated in brightness (i.e., the original design) and the rest, while having the same overall design, are thought to have visors.
The addition of such visors is a voluntary action by SpaceX in response to the initial outcry from the astronomical community. 
Early attempts to characterize this retrofitted mitigation showed some promise in the reduction of light pollution \citep[e.g.,][]{Mallama2021}, but as we will show, there is significant variability in satellite brightness. For ease of discussion, we distinguish between the original Starlink design and the visorsat addition as Starlink and Starlink-V, respectively. 

The target satellites and select observing details are listed in Table \ref{tbl:satlist}.
We, unfortunately, cannot confirm that the satellites indicated as Starlink-Vs do indeed have visors or that the visors are correctly deployed. 
Starlink-1436 is the first visorsat \citep[e.g., see discussion in][]{tregloan-reed2021}, placed on orbit as a test case and widely discussed on social media.
Additional Starlink satellites were launched without visors until the tenth Starlink mission, which according to SpaceX \citep{spacexVisorNews}, included deployable visors for all 57 Starlink satellites on board the 6 August 2020 Falcon 9 launch.
Subsequent mission summaries on SpaceX's website do not clarify whether the satellites include visors. 
However, SpaceX personnel have stated publicly that all satellites after the August launch would indeed be equipped with visors\footnote{see, e.g., the summary at \url{https://directory.eoportal.org/web/eoportal/satellite-missions/s/starlink}.}. 
Public satellite information does not confirm either way, other than referencing some public statements.
Nonetheless, if we take the available information at face value, then satellites Starlink-1522\footnote{This is the lowest Starlink number associated with the August launch, according to the satellite catalogue, available at celestrak.com.} onward, as well as Starlink-1436, have deployable visors.

\begin{table}[]
    \centering
        \caption{Target Information}
    \begin{tabular}{l c c c c c c c c }

 Name & UTC\tablenotemark{a} & RA(J2000) & $\delta$(J2000) & Azimuth & Elevation  & Rate  & Range  & STO \\
    & & & & deg & deg &   $\arcsec/s$ & km & deg\\
    \hline
  STARLINK-2077(v)\tablenotemark{b} & 05:45:11 &  15 28 26.31  & +08 53 52.7 & 216.00 & 45.16 & 2021 & 749.0 & 69\\
 STARLINK-1392 & 05:57:29 &  15 52 24.66  & +13 22 50.4 & 214.71 & 50.41 & 2171 & 697.1 & 66\\
 STARLINK-1747(v) & 06:13:47 &  18 57 30.34  & +16 29 47.3 & 145.59 & 53.97 & 2263 & 668.3 & 39\\
 STARLINK-1728(v) & 06:28:56 &  18 45 17.69  & +27 54 54.4 & 148.44 & 66.96 & 2537 & 595.5 & 51\\
 STARLINK-1355 & 06:41:17 &  17 42 31.87  & +40 37 24.5 & 207.97 & 81.21 & 2705 & 558.0 & 68\\
 STARLINK-1300 & 06:55:00 &  19 35 38.38  & +17 45 39.1 & 145.94 & 55.45 & 2300 & 657.5 & 39\\
 STARLINK-2476(v) & 07:07:22 &  19 59 51.14  & +12 51 33.2 & 144.76 & 49.83 & 2154 & 702.5 & 34\\
 STARLINK-2565(v) & 07:11:25 &  18 35 44.49  & +48 04 10.5 & 154.91 & 89.52 & 2734 & 551.7 & 71\\
 STARLINK-1529(v) & 07:28:02 &  18 13 43.49  & +34 24 40.8 & 209.57 & 74.29 & 2642 & 571.7 & 60\\
 STARLINK-2530(v) & 07:43:20 &  17 26 41.52  & +08 48 58.9 & 216.05 & 45.12 & 2022 & 748.8 & 45\\
 STARLINK-1549(v) & 07:55:38 &  17 51 22.27  & +13 20 05.0 & 214.55 & 50.47 & 2172 & 696.6 & 44\\
 STARLINK-1012 & 08:11:54 &  20 55 40.42  & +16 21 13.5 & 145.74 & 53.91 & 2263 & 668.6 & 42\\
 STARLINK-1009 & 08:28:28 &  20 42 08.85  & +28 54 50.0 & 148.93 & 68.18 & 2558 & 590.7 & 52\\
 STARLINK-1498 & 08:42:05 &  21 41 52.96  & +10 30 13.8 & 143.73 & 47.02 & 2076 & 729.3 & 43\\
 STARLINK-1561(v) & 08:55:55 &  19 24 09.88  & +26 48 08.4 & 211.13 & 65.84 & 2518 & 600.0 & 48\\
 STARLINK-1576(v) & 09:09:34 &  20 32 40.24  & +48 16 44.6 & 192.59 & 89.83 & 2733 & 552.1 & 71\\
 STARLINK-1037 & 09:24:58 &  21 53 23.21  & +23 00 29.2 & 147.19 & 61.48 & 2460 & 614.7 & 55\\
 STARLINK-1060 & 09:39:47 &  22 07 49.16  & +22 57 10.8 & 147.42 & 61.47 & 2436 & 620.7 & 57\\
 STARLINK-2063(v) & 09:55:00 &  20 23 29.08  & +26 28 37.0 & 210.80 & 65.58 & 2518 & 600.2 & 49\\
 STARLINK-1464 & 10:10:00 &  22 54 30.14  & +16 16 48.5 & 145.60 & 53.83 & 2261 & 669.2 & 60\\
 STARLINK-2252(v) & 10:25:17 &  21 12 44.06  & +34 10 07.6 & 208.36 & 74.25 & 2643 & 571.5 & 59\\
 STARLINK-2249(v) & 10:41:41 &  20 56 12.20  & +21 04 56.1 & 212.77 & 59.38 & 2396 & 631.2 & 46\\
 STARLINK-2195(v) & 10:53:10 &  22 51 39.83  & +35 54 22.5 & 149.43 & 75.93 & 2598 & 580.8 & 73\\
 \hline
    \end{tabular}
    \label{tbl:satlist}
    \tablecomments{Positional information and range as provided by the JPL Horizons ephemeris service.  }
    \tablenotetext{a}{UTC times represent the expected passage of the satellite through the centre of the FOV}
    \tablenotetext{b}{Visorsat Starlink satellites are denoted with a '(v)'}
    \tablenotetext{c}{Solar-Target-Observer (STO) angle (essentially the solar phase angle)}

\end{table}

\section{Observations and Measurements}

All observations were taken using a $g'$ filter with 30 second exposures. 
Each integration began 15 seconds prior to the central UTC time listed in Table \ref{tbl:satlist} to ensure that the satellite would be observed as it passed through the Plaskett's $22'\times 11'$ (edge-to-edge on the CCD) FOV.

Images were processed in two ways to check for consistency. The first method used the Spaceguard pipeline \citep{SpaceguardPipeLine}, which includes bias subtraction, flat-fielding, image flattening, and filtering for the removal of image artefacts (e.g., hot pixels, cosmic rays, etc.). 
For each image, source catalogues are generated of all stars in the field using Gaussian fitting. 
One of a selection
of astrometric catalogues are then queried (Tycho2, PPM, Carlsberg Meridian Circle, USNO, and 2MASS)
and predictions of the detector positions of each catalogue source are calculated using the
world coordinate system (WCS) information in the image header as a first pass.
The two lists are then
cross-correlated, the catalogue source positions are re-calculated using Gaussian fitting, and
the final WCS is determined. 
All pixels are then re-sampled to tangential projections with constant
scale ($0\arcsec .62$) in the ICRS reference frame. 
All re-sampling maintained flux conservation for each image.

The second method used astropy and python scripts to bias subtract, flat-field, flatten, and median filter the images. 
The flattening was done by subtracting from a given pixel the median value as determined from 128x128 pixel grid. 
The large grid was needed to avoid forming peak-and-valley structure in the images along the streaks and the brightest stars. To speed up this process, a subgrid consisting of 20x20 pixel bins was used, with the median subtraction done on all subgrid pixels using the large window centred on the subgrid. 
Each image was then astrometrically calibrated using the astrometry.net software package \citep{astrometry}.

With the astrometrically calibrated images, bright stars were selected that sample  the relevant image area (i.e., in the vicinity of the satellite streaks) and, to the extent possible, were free of contamination from other stars or image artefacts. 
Some fields were very crowded, so the selected stars were often not the brightest stars in the field. 
Fortunately, the 30 second integration times used on each image provided plenty of very high signal-to-noise ratio (SNR) stars for calibration. 

The selected stars were cross-referenced with the \Gaia~catalogue, and their $g$ magnitudes were obtained by using the \Gaia~$G$, $G_{BP}$, and $G_{RP}$ photometry with the SDSS12-\Gaia~photometric transformation fits for GEDR3, with a reported uncertainty of $\sigma_{trans}=0.075$ mag.\footnote{Available at: \url{https://gea.esac.esa.int/archive/documentation/GEDR3/Data_processing/chap_cu5pho/cu5pho_sec_photSystem/cu5pho_ssec_photRelations.html}}
The two different methods for astrometric calibration were self-consistent with source identification and cross referencing with \Gaia\ positions. 

The transformed \Gaia~magnitudes, $m_g^G$, were then compared with the corresponding instrumental magnitudes, $m_{inst}$, as determined through aperture photometry. 
The zero point for each star was set according to $m_0 = m_g^G-m_{inst}$.  
The resulting distribution of zero points were clipped to remove outliers.
The $m_0$'s for an image, which included a minimum of five stars after clipping, were averaged to determine the image's zero point.
The clip-and-average results were typically within about 0.01 mag of the median value for the zero points. 

We note that we use a $g'$ filter, but have calibrated the photometry using $g$ magnitudes. 
The difference between the two bands is small \citep[$\lesssim 0.01$;][]{tucker2006}, much smaller than the uncertainty $\sigma_{trans}$ in the transformation from \Gaia\ bands to $g$.
We therefore report the magnitudes as $g$ band to reflect the calibration.

In determining the satellite magnitudes, there are several complications that are introduced by the wait-and-catch method, all of which were anticipated: (1) A satellite remains in the FOV for only a small fraction of the 30-sec exposure; (2) the streaks cut across multiple stars and image artefacts; and (3), the streaks could, in principle, vary in brightness across the image. Fortunately, we did not see large streak brightness variations, although there were measured variations on the level of a few tenths of a mag in a few cases. 

To address the second issue, we did not use the entire streak. 
Rather, we selected as long of a section as practicable that was free of substantial contamination.
In a few cases, several regions were selected and combined.
Altogether, the shortest streak length that we used was $62\arcsec$, while the longest was about $370\arcsec$. 
The lowest SNR (just taken to be Poisson dominated) is 120, which corresponded to the dimmest satellite in the sample. 
The typical SNR was a few hundred. 
The total flux inside a rectangle region along the selected section of the streak was used for the photometric measurements, with adjacent regions used to determine the background for subtraction. 

The extrapolated magnitude for each satellite is then calculated using
\begin{equation}
    m_g = \left<m_0\right> + m_{inst} -2.5 \log_{10} \left( \frac{R}{L} t_{EXP}\right),
\end{equation}
where $R$ is the sky rate of the satellite as determined from the JPL ephemerides, $L$ is the measured streak length, and $t_{EXP}$ is the exposure time, which again is 30 seconds for these observations.

Finally, we define an `absolute magnitude', $H_g^{550}$, for the satellites as the brightness the satellite would be if seen from 550 km. 
We note that different papers use a different standard for this, but we use 550 km to reflect the orbital altitude for the satellite shell in question. To be verbose, 
\begin{equation}
    H_g^{550} = m_g - 5 \log_{10} \left( \frac{r}{550~\rm{km}}\right)
\end{equation}

\section{Results}

The measurements are presented in Table \ref{satmags}. 
The apparent brightness distribution  varies between $m_g=4.7$ mag and 10.0 mag, with the brightest being a Starlink-V and the dimmest a Starlink satellite.  

\begin{figure}
    \centering
    \includegraphics[width=0.45\textwidth]{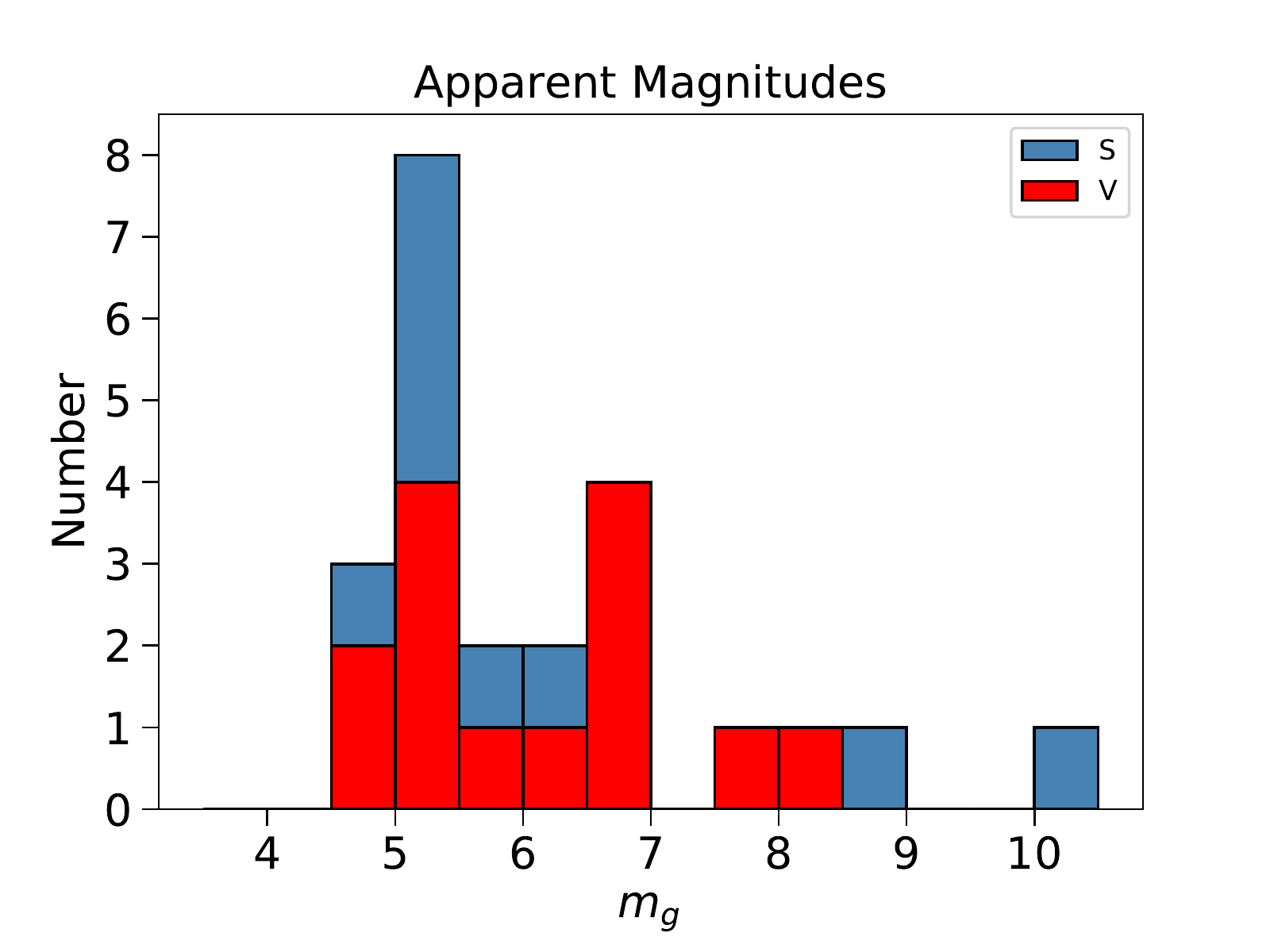}
    \includegraphics[width=0.45\textwidth]{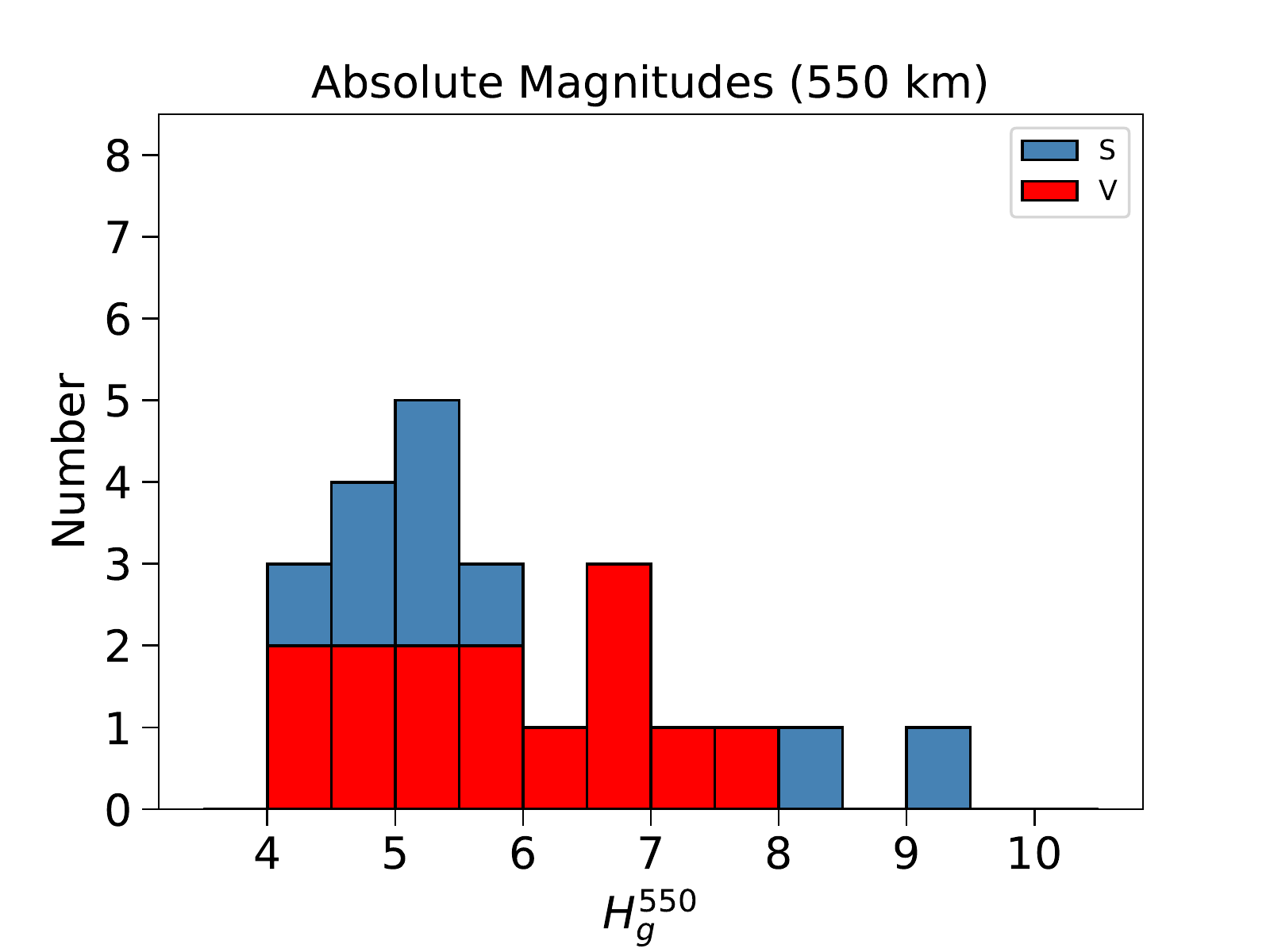}
    \caption{Magnitude distributions for the Starlink satellites observed in this study. Each satellite was observed once only. Left: apparent $m_g$ magnitudes, with blue showing the contribution to each bin from Starlinks and red the Starlink-Vs. Right: similar to the left panel, but for the absolute magnitude, as defined by seeing the satellite at a range of 550 km. No STO correction is applied. Because some values fall on bin edges, we use the interval $[a,b)$ for edges $a$ and $b$ when assigning a magnitude to a bin.}
    \label{fig:histos}
\end{figure}

Table 2 is visualized in Figure \ref{fig:histos}. 
The median for the apparent magnitudes is $\overline{m}_g=5.7$ mag, while it is $\overline{H}_g^{550}=5.3$ mag for the absolute  distribution. Their respective standard deviations are 1.4  mag and 1.3  mag.  
This includes the general trend of the satellites with STO, which is discussed further below. When the trend is removed, the magnitude variation is 0.5 mag for the apparent magnitudes.
Isolating just the Starlink-V and Starlink distributions, the apparent magnitude medians are $\overline{m}_g({\rm visor})=6.1 $ mag and $\overline{m}_g({\rm no~visor})=5.2$ mag, respectively. 
The absolute magnitudes are likewise $\overline{H}_g^{550}({\rm visor})=5.7$ mag and $\overline{H}_g^{550}({\rm no~visor})=5.1$ mag.   
Thus, the Starlink-V mitigation efforts through visors do have a measurable effect, with an overall dimming by $\Delta H_g^{550}=0.6$ mag.  
However, the medians of both distributions are still bright and are naked-eye visible. Moreover, the substantial spread further suggests that mitigation needs to be part of the design process, and that retrofits may have limited impacts as they can still be bright a large fraction of time. 

\begin{table}[]
    \centering
        \caption{Satellite Brightness}
    \begin{tabular}{l l c c }

 Number & Name & Apparent Magnitude ($m_g$) & Absolute Magnitude ($H_g^{550}$)\tablenotemark{a} \\
\hline
1 & STARLINK-2077(v) & 6.5 & 5.8\\
2 & STARLINK-1392 & 5.7 & 5.2\\
3 & STARLINK-1747(v) & 5.3 & 4.8\\
4 & STARLINK-1728(v) & 5.4 & 5.2\\
5 & STARLINK-1355 & 5.2 & 5.1\\
6 & STARLINK-1300 & 4.8 & 4.4\\
7 & STARLINK-2476(v) & 4.9 & 4.3\\
8 & STARLINK-2565(v) & 6.5 & 6.5\\
9 & STARLINK-1529(v) & 6.4 & 6.3\\
10 & STARLINK-2530(v) & 7.9 & 7.2\\
11 & STARLINK-1549(v) & 8.2 & 7.7\\
12 & STARLINK-1012 & 8.5 & 8.1\\
13 & STARLINK-1009 & 5.2 & 5.0\\
14 & STARLINK-1498 & 10.0 & 9.4\\
15 & STARLINK-1561(v) & 5.7 & 5.5\\
16 & STARLINK-1576(v) & 6.8 & 6.8\\
17 & STARLINK-1037\tablenotemark{b} & 6.2 & 5.9\\
18 & STARLINK-1060 & 5.2 & 4.9\\
19 & STARLINK-2063(v) & 5.0 & 4.8\\
20 & STARLINK-1464 & 5.1 & 4.7\\
21 & STARLINK-2252(v) & 5.4 & 5.3\\
22 & STARLINK-2249(v) & 4.7 & 4.4\\
23 & STARLINK-2195(v) & 6.9 & 6.7\\
    \end{tabular}
    \label{tbl:satmags}
    \tablecomments{All measurements are in the $g'$ filter. Uncertainties estimated to be about 0.1 mag. Row numbers are for comparisons with figures.}
    \tablenotetext{a}{As seen at 550 km, but without a STO (phase angle) correction.}
    \tablenotetext{b}{Inactive}\label{satmags}

\end{table}

\begin{figure}
    \centering
    \includegraphics[width=0.45\textwidth]{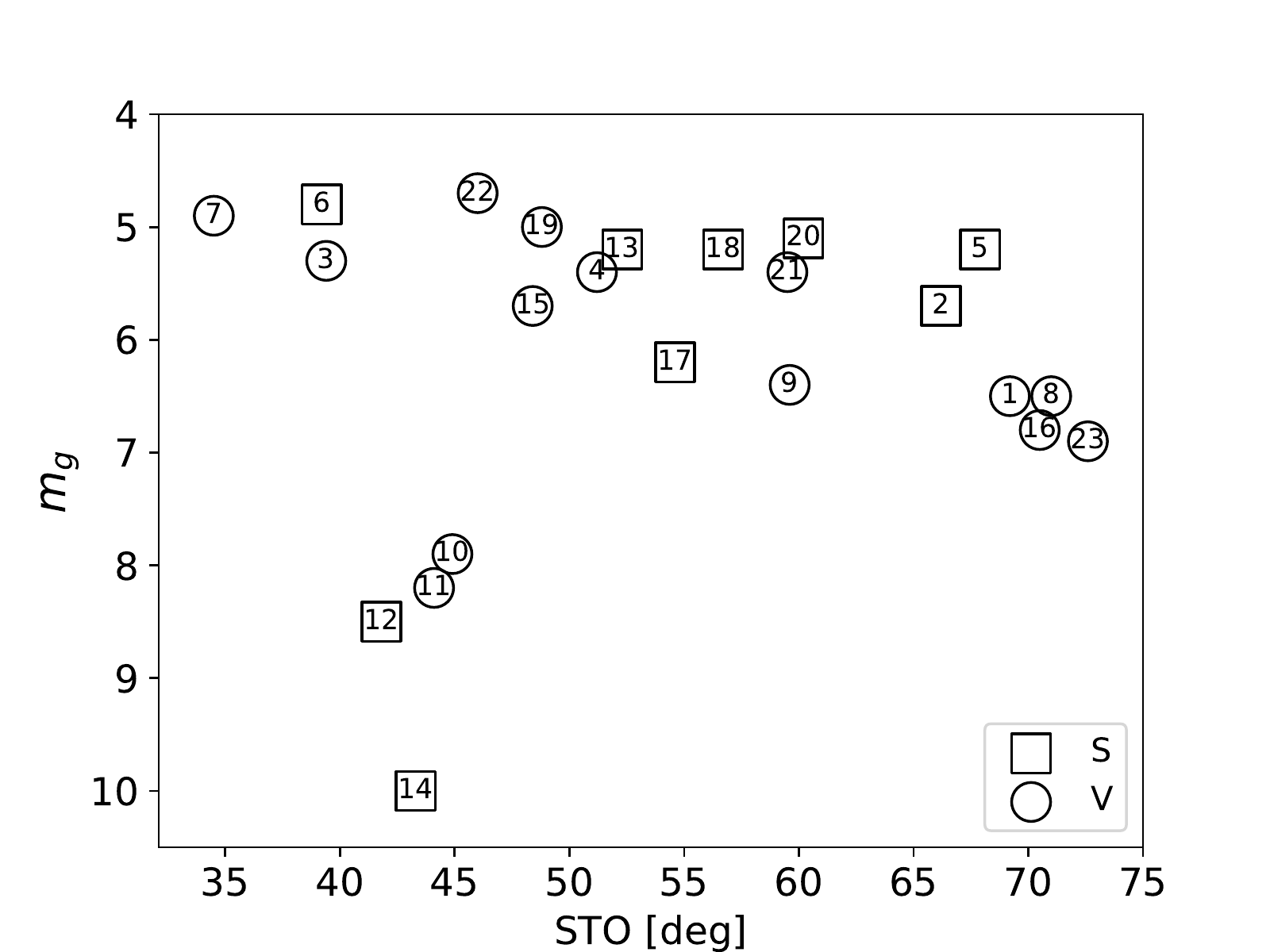}
    \includegraphics[width=0.45\textwidth]{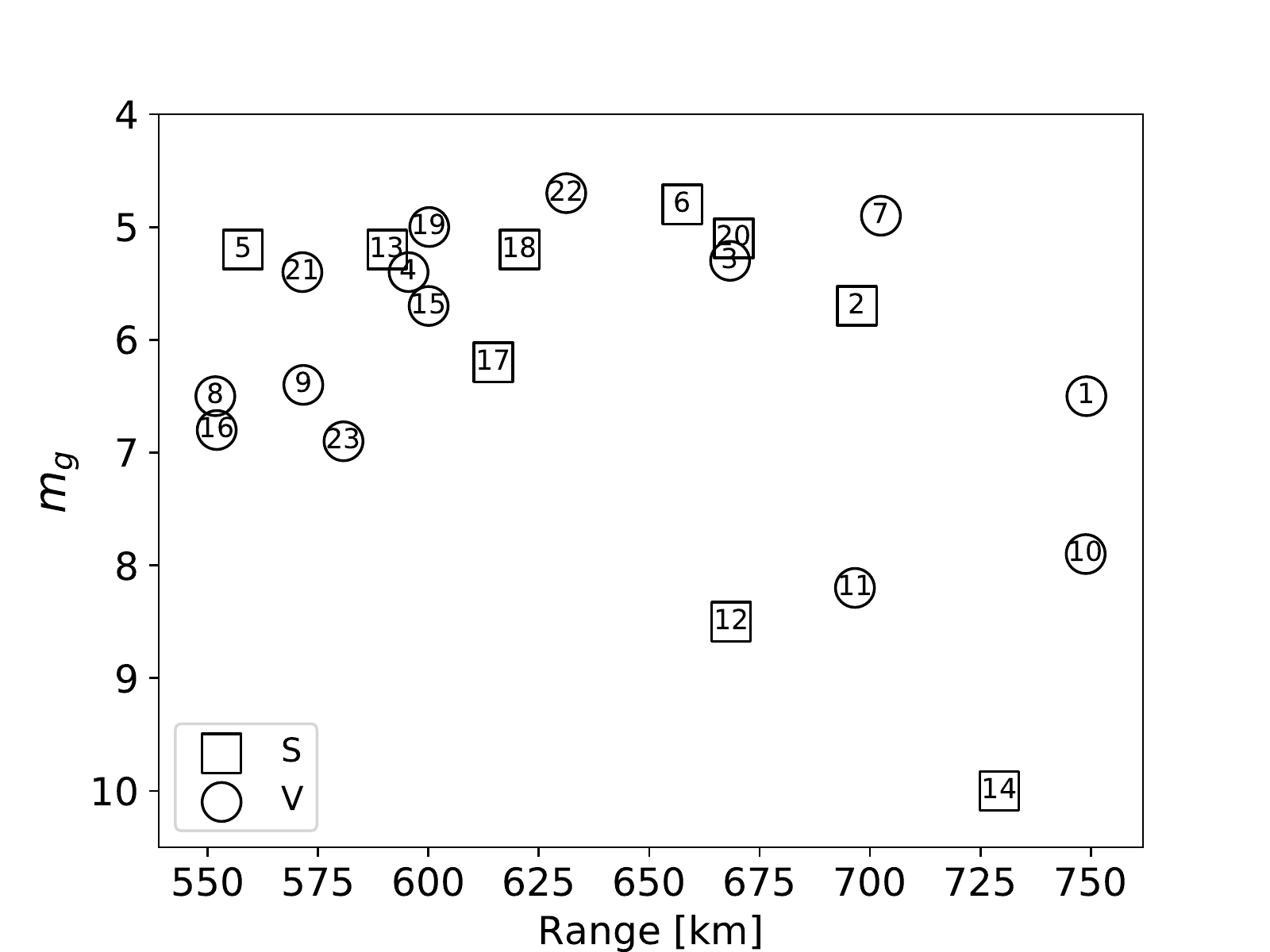}
    
    \includegraphics[width=0.45\textwidth]{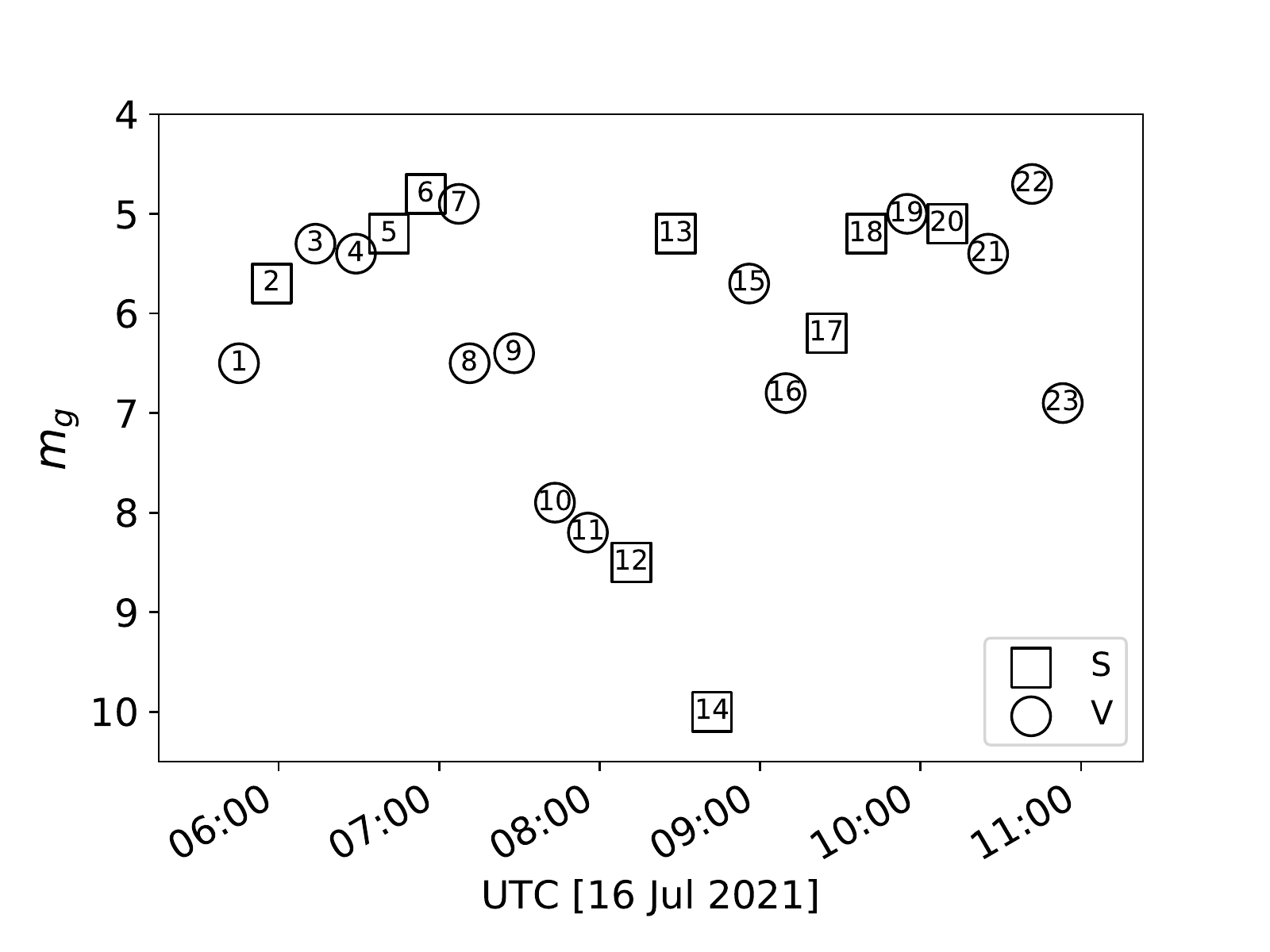}
    \includegraphics[width=0.45\textwidth]{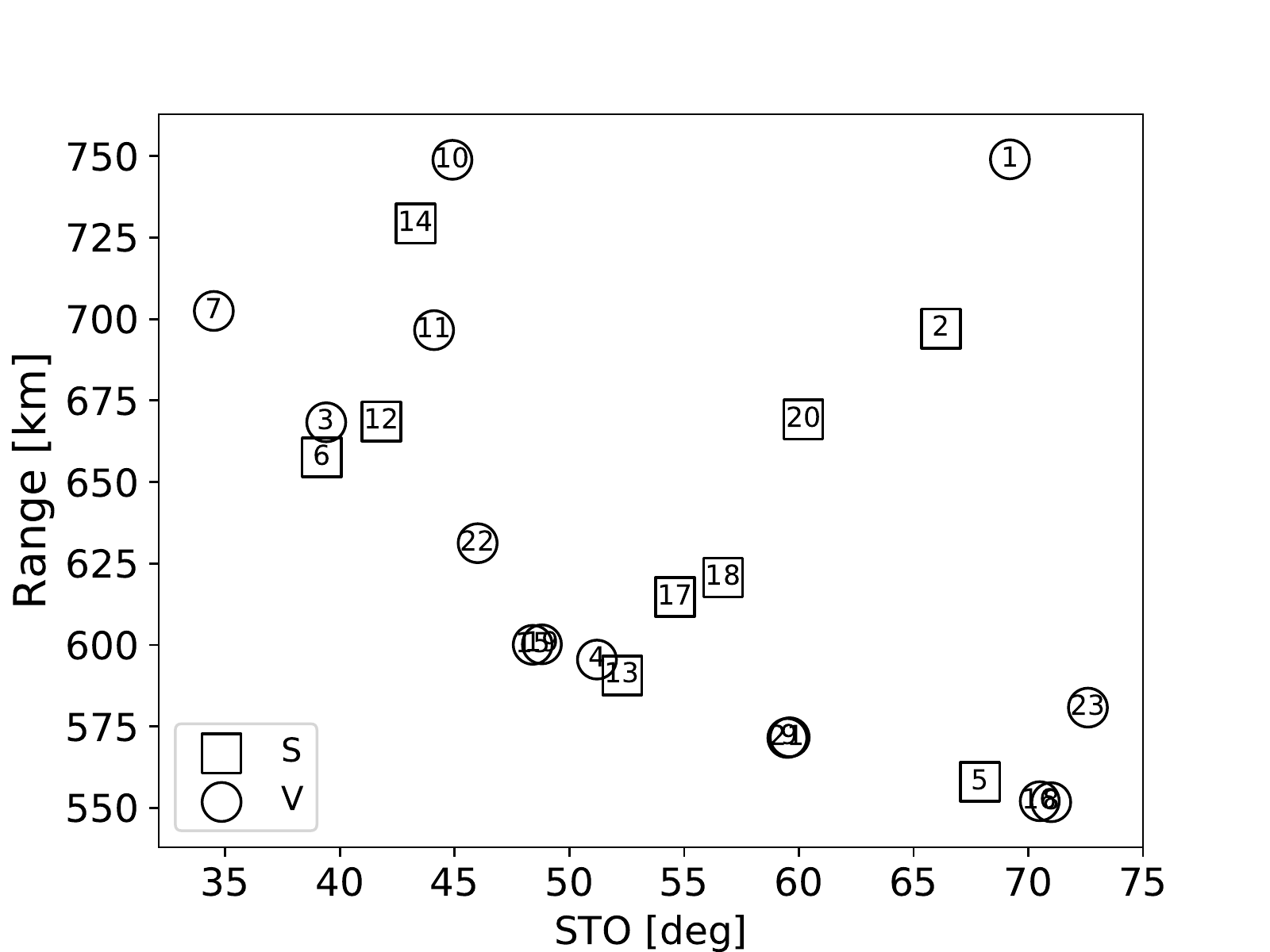}
    \caption{ Select observational properties for the observed satellites, emphasizing Starlinks (square) Starlink-Vs (circle). A slight dimming of the satellites is seen with increasing STO. A sudden change in brightness occurs around $43^\circ$ for both Starlink and Starlink-V.
    This feature is not understood. 
    The brightest satellites are seen at a range near 650 km due to the combination of distance and favourable STO. 
    The satellites with the smallest possible range (550 km), due to the orbital altitude, occur near an STO of about $70^\circ$. 
    Higher STOs were not observed due to restrictions on telescope azimuth for slew considerations.}
    \label{fig:distros}
\end{figure}

The satellite distributions can further be shown against their STO, range, and time (Fig.~\ref{fig:distros}).
There is a general trend of dimming with increasing STO.
This behaviour is non-trivial in that a satellite passing near the observer's zenith will have an STO that approaches $90^\circ$, but adjusted for the Sun's angular distance below the horizon. This is also when the satellite is closest to the observer, which by itself would tend to make the satellite brighter. 
However, a shallower STO is able to compensate for the difference in range, up to a point, which places the shallower STOs here among the brightest observed satellites.
Of course, the STO is not unique to a given range, which can be seen in Figure \ref{fig:distros}. 
For the restrictions placed on our observing elevation and azimuth, the brightest satellites occur at a range of about 650 km. 
These same restrictions set the limits on the range of STOs observed.

A surprising feature in the data is the dimming of the satellites at an STO of about $43^\circ$. This dimming cannot be explained by range. The time of observation also does not show clear evidence alone for this feature given that the dimmest two satellites are separated by one of the brightest.
Moreover, there is not a significant change in the image zero points to suggest that there is a local sky phenomenon that is compromising some of the measurements.

\section{Discussion}

Characterizing the brightness distribution of satellites, their variability, and dependence on   STO (solar phase angle) is necessary for assessing the degree to which satellites will interfere with astronomy and stargazing. It is further required for modelling the impacts of proposed satellite systems \citep[e.g.,][]{lawlerModels}, as well as monitoring mitigation efforts and, to the extent practicable, identifying possible regulation and industry best practices that would have the greatest impact on reducing light pollution while allowing the development of orbital infrastructure.  

The observations here show that the satellites have significant scatter in brightness, which is not improved when correcting to a range of 550 km. 
This is not surprising given the range of STOs among the satellites. 
We thus can consider different phase models, with the simplest being the diffuse sphere \citep{pradhan_2019}.
Specifically, the model $g$ magnitude is
\begin{equation}
   m_g = -26.47 - 2.5 \log_{10} \left\{ \frac{2}{3\pi^{p+1}} \zeta \left[ \left( \pi-{\rm STO}\right) \cos\left({\rm STO}\right)+\sin\left({\rm STO}\right) \right]^p\right\}  + 5 \log_{10}\left(R_{\rm m}\right).\label{eqn:phasepower}
\end{equation}
where $\zeta$ is the effective area of the satellite in square metres (i.e., the product of the albedo and cross-sectional area), $R_{\rm m}$ is the range in metres, and the index $p=1$. 

\begin{figure}
    \centering
    \includegraphics[width=0.45\textwidth]{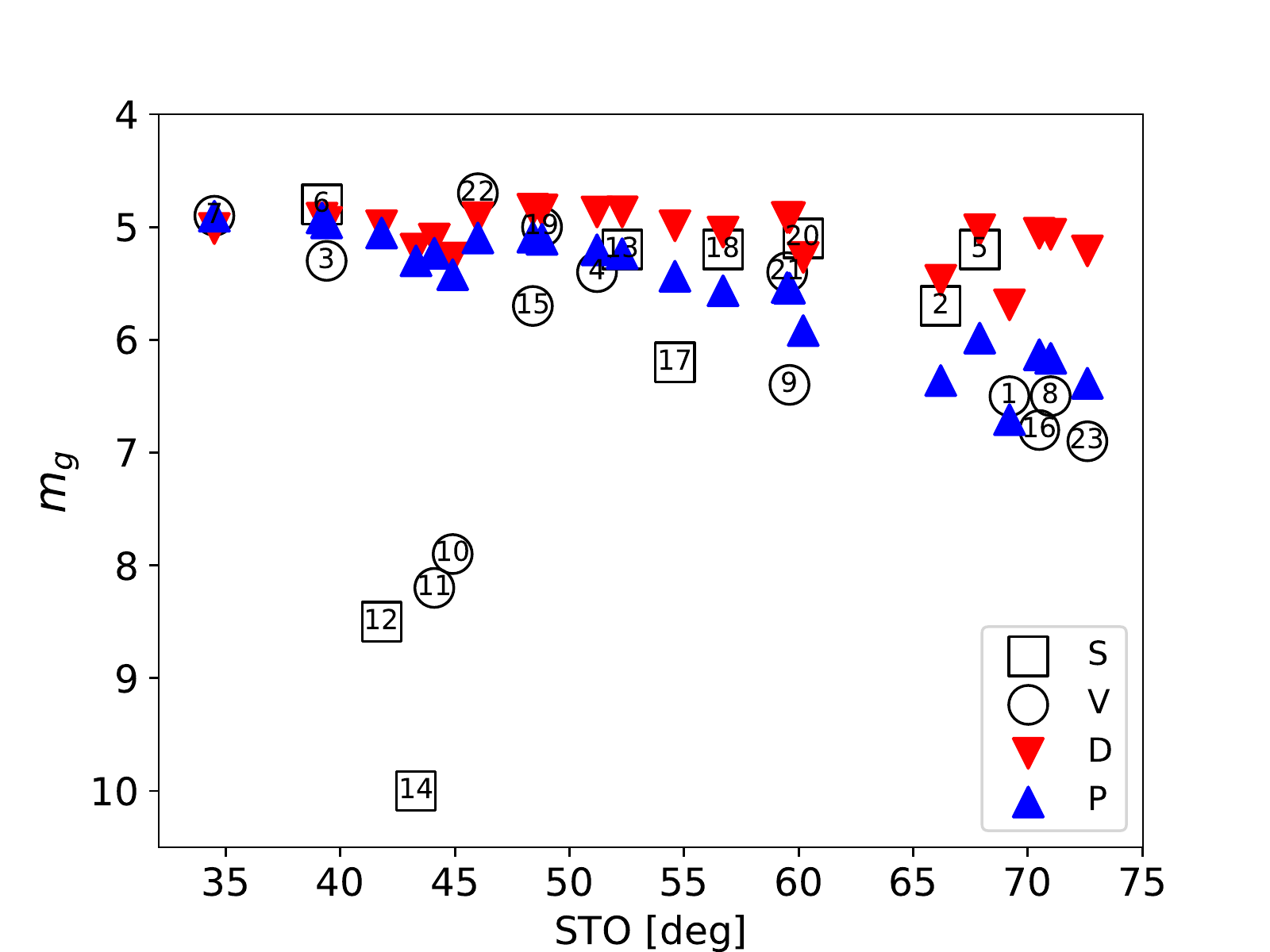}\includegraphics[width=0.45\textwidth]{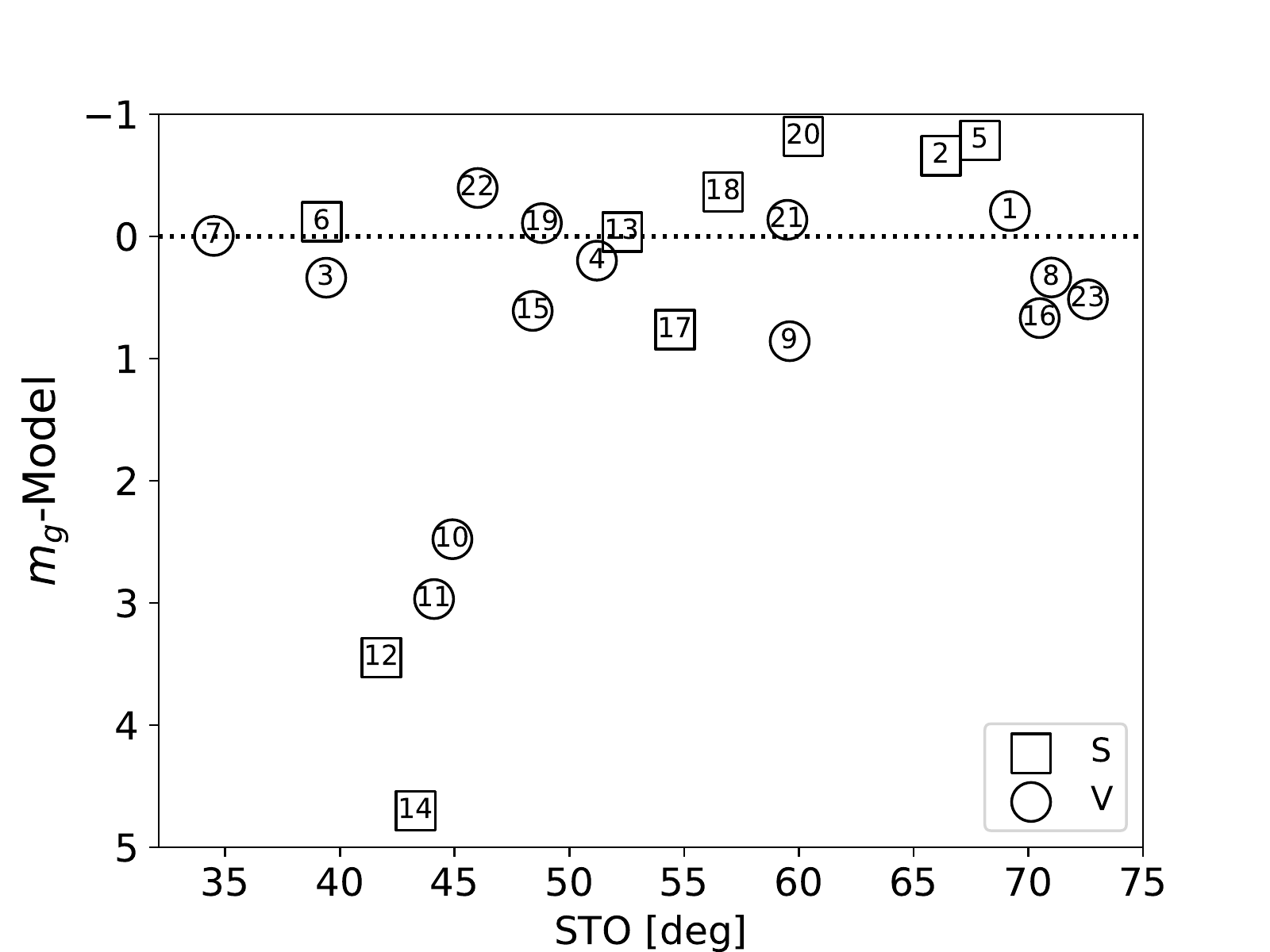}
    \caption{Similar to Figure 2, but including a diffuse sphere model for the apparent magnitude (red, downward pointing triangle) and a heuristic power law STO (phase) model (blue, upward pointing triangle). The right panel shows the difference between the observed magnitudes and the predictions from the heuristic phase model. }
    \label{fig:diffuseSphere}
\end{figure}

The result is shown in Figure \ref{fig:diffuseSphere}.
We use $\zeta = 0.7\rm~m^2$, which provides a reasonable match to the low STO satellites, but does not dim fast enough for higher STOs. 
Noticing that a faster dependency is required, we explore possible variations of the diffuse sphere model in which $p\ne 1$.

A Monte Carlo Markov Chain routine is written in python using a Metropolis-Hastings algorithm with a Gibbs sampler (sufficient for this application), using uniform priors to explore possible $p$ and $\zeta$. 
We assume the data have uncertainties of $0.1$ mag and exclude the four peculiarly faint satellites at ${\rm STO}\approx43^\circ$.
The result is $\zeta\approx 1.1\rm~m^2$ and $p\approx3.1$. 
We hesitate to place uncertainties on these very approximate relations.  
The results are nonetheless shown in Figure \ref{fig:diffuseSphere}.

The heuristic model magnitudes are subtracted from the data (Fig.~\ref{fig:diffuseSphere}), resulting in a dispersion of 0.5 mag among the residuals (excluding again the four dimmest satellites). We caution against reading too much into this model at this time, as additional data are needed to test it. 
Moreover, we only cover a small range of STOs. 

With this in mind, an additional behaviour can be noted. First, the Starlink-Vs appear to fall off in brightness slightly more rapidly with STO than the Starlink satellites. 
This is consistent with the differences in the median magnitudes for the two populations, but also shows that so far the effectiveness of Starlink-Vs is highly variable and does not help with some of the brightest orientations. 

Recent work by \cite{mallama2021b} combines about 10,000 MMT-9 robotic observations of Starlink-Vs using a clear filter. 
They report that the magnitudes are within about 0.1 mag of the Johnson V band. 
The median of their Starlink-V observations corrected to 550 km is 5.5 $V_{\rm MMT}$ mag.\footnote{The subscript is used to denote that the observations are based in the clear MMT-9 filter and transformed to an approximate $V$ band.}
When we take into account the $g-V=0.3$ colour of the Sun, assuming no wavelength dependence for the albedo, their median $\overline{H}_g^{MMT}= 5.8$ mag compared with our $\overline{H}_g^{550}=5.7$ mag, consistent to within uncertainties. 

\cite{mallama2021b} also find that a quadratic fit to the STO-brightness dependence can explain their observations. 
We note that the quadratic fit behaves nearly identically to a diffuse Lambertian sphere over the range shown in Figure 3 (this work), but with the leading term set to 3 instead of their nominal value. 
That correspondence means our data exhibits a slightly different dependence on STO than the MMT-9 data. 
The origin of this is unclear, and we refrain from discussing it further other than to note that because the satellites are not actually spheres, the detailed viewing angle will matter. 
Regardless, this study and \cite{mallama2021b} find an inherent satellite magnitude dispersion of about 0.5 mag after subtracting their respective STO trend. 
We conclude that this is an inherent dispersion due to the shape of the satellite and that the visors do not prevent this from occurring.

The results can be used in modelling the impacts of satellites on the night sky, as done in \cite{lawlerModels}, at least for Starlink-like satellites. 
As additional photometric data are acquired in astrophysical bands, we will be able to ensure higher fidelity between observations and models, at least in a statistical sense. 

The scatter demonstrates that analysis of mitigation efforts must be done statistically, with the need to consider observations that span a wide range of STOs and orientations of the spacecraft relative to the observer. 

These observations only measure the brightness of Starlink satellites in a single filter. Additional observations in different bandpasses are needed, as well as measurements of different satellite designs (e.g., OneWeb). 
As part of seeking cooperation with satcon operators toward minimizing the impact that their satellites have on the sky, we need to be vigilant with independent verification of their mitigation efforts. 
We also need to consider assessments of single satellite magnitudes, while investigating cumulative impacts when practicable. 
Astronomy is a fundamental way to explore space, test our understanding of physical laws, and detect impact hazards, among many other things -- interference with astronomy has wide ramifications for science, education, and safety.

The observational planning scripts, calibrated images, associated photometry, and analysis scripts used for this paper are available at \url{https://github.com/norabolig/Boleyetal-Plaskett-2021}. 
The astropy simple reduction tools are available at \url{https://github.com/norabolig/ABIRL}.

\begin{acknowledgments}
 The University of British Columbia is situated on the traditional, ancestral, and unceded territory of the Musqueam people.  The University of Regina is located on Canadian Treaty 4 land, which is the traditional territories of the n\^{e}hiyawak, Anih\v{s}in\={a}p\={e}k, Dakota, Lakota, and Nakoda, and the homeland of the M\'{e}tis/Michif Nation.

This work is based on observations obtained at the Dominion Astrophysical Observatory, NRC Herzberg, Programs in Astronomy and Astrophysics, National Research Council of Canada. It also used the facilities of the Canadian Astronomy Data Centre operated by the National Research Council of Canada with the support of the Canadian Space Agency.

This research was supported in part by NSERC Discovery Grants RGPIN-2020-04635 (ACB) and  RGPIN-2020-04111 (SML), as well as the Canada Research Chairs program.

\end{acknowledgments}

\software{This research was made possible by the open-source projects \texttt{astrometry.net} \citep{astrometry}, \texttt{astropy} \citep{astropy}, \texttt{python} \citep{python3}, \texttt{numpy} \citep{numpy},  and \texttt{matplotlib} \citep{matplotlib, matplotlib2}.}

\bibliography{bib}{}
\bibliographystyle{aasjournal}

\end{document}